\documentclass[twocolumn,twoside,slac_two]{revtex4}
\usepackage{graphicx}
\usepackage{fancyhdr}
\usepackage{natbib}

\usepackage{graphicx}
\usepackage{color}
\usepackage{amsmath,amssymb,amsfonts,dsfont}
\usepackage{multirow}
\usepackage{sistyle}

\pdfoutput=1 

\newcommand{\sigmav}{$\langle\sigma v\rangle$ }

\def\lsim{\mathrel{\rlap{\lower4pt\hbox{\hskip1pt$\sim$}}
    \raise1pt\hbox{$<$}}}                
\def\gsim{\mathrel{\rlap{\lower4pt\hbox{\hskip1pt$\sim$}}
    \raise1pt\hbox{$>$}}}                

\begin{document}


\title{\boldmath Isotropic diffuse $\gamma$-ray background: unveiling Dark Matter components beyond the contribution of astrophysical sources.}

\thispagestyle{fancy}


\author{Mattia Di Mauro,}\email{mattia.dimauro@to.infn.it}
\affiliation{Dipartimento di Fisica, Universit\`a di Torino, via P. Giuria 1, 10125 Torino, Italy}
\affiliation{Istituto Nazionale di Fisica Nucleare, Sezione di Torino, Via P. Giuria 1, 10125 Torino, Italy}
\affiliation{LAPTh, Universit\'e de Savoie, CNRS, 9 Chemin de Bellevue, B.P.\ 110, F-74941 Annecy-le-Vieux, France}

\begin{abstract}
We present the first interpretation of the new isotropic diffuse $\gamma$-ray background (IGRB), measured by the {\it Fermi} Large Area Telescope (LAT), based on a statistical analysis.
We demonstrate that the $\gamma$-ray emission from unresolved active galactic nuclei and star forming galaxies is consistent with the {\it Fermi}-LAT IGRB data within the uncertainties both on the choice of the Galactic diffuse emission model and on the $\gamma$-ray emission mechanism of these sources.
Furthermore, adding to the extragalactic sources the contribution from a smooth Galactic halo of annihilating weakly interacting dark matter (DM) particles, we are able to set stringent limits on the DM annihilation cross section.
Finally, we demonstrate that the addition of DM can significantly improve the fit to IGRB data.
\end{abstract}

\maketitle

\section{Introduction}
\label{sec:introcomp}
Recently, the Large Area Telescope (LAT) on board the {\it Fermi Gamma-ray Space Telescope (Fermi)} 
has published a new measurement of the isotropic diffuse $\gamma$-ray background (IGRB) and the extragalactic $\gamma$-ray background (EGB) in the energy range 100 MeV-820 GeV \cite{igrb_2014}.  
The origin of this $\gamma$-ray residual represents one of the most  mysterious open problems in astrophysics. 
The IGRB is usually associated to the $\gamma$-ray emission from unresolved, namely not detected by the {\it Fermi}-LAT, blazars, misaligned active galactic nuclei (MAGN) and star forming (SF) galaxies \cite{DiMauro:2013zfa,DiMauro:2013xta,Ackermann:2012vca}. 
The most powerful Galactic contributors of the IGRB are expected to be pulsars, due to the large sample of detected sources. 
However, very recently in Ref. \cite{Calore:2014oga} the $\gamma$-ray emission from high-latitude ($|b|>20^{\circ}$) unresolved pulsars has been derived to account for, at most, the $1\%$ of the IGRB.
\\
The annihilation of dark matter (DM) particles, in the halo of the Milky Way (MW) and in external galaxies, constitutes a possible exotic mechanism for the production of $\gamma$ rays.
Indeed, one of the most promising strategies for the search of DM, in the scenario of weakly interacting massive particles (WIMPs), is the indirect detection through $\gamma$ rays produced from the annihilation of WIMPs.
\\
We first explore in Sec.~\ref{sec:astro} at which extent the astrophysical populations may explain the EGB and IGRB data and then in Sec.~\ref{sect:dm} we add the contribution from a DM Galactic halo to constrain the DM annihilation cross section. 
We use a statistical fitting procedure which includes both the {\it Fermi}-LAT data errors and the theoretical uncertainties on the $\gamma$-ray emission from astrophysical sources.
For the first time we show the effect of the choice of the Galactic diffuse emission (GDE) model, used to derive the IGRB and EGB data, and how this affects the results on DM.

\section{{\it Fermi}-LAT IGRB data explained with astrophysical sources}
\label{sec:astro}
At least the 10\% of the $\gamma$-ray photons, detected at latitude $|b|>20^{\circ}$ by the {\it Fermi}-LAT, are emitted from Galactic and extragalactic resolved sources.
Indeed, the {\it Fermi}-LAT catalogs contain thousands of point sources.
One of the most natural explanation for the origin of the IGRB, is that it arises from the superposition of a numerous population of unresolved sources with a flux lower than the point source sensitivity threshold of the LAT \cite{DiMauro:2013zfa,DiMauro:2014wha}.
\\
A diffuse $\gamma$-ray flux has been predicted for various source populations. 
A large fraction of the IGRB is expected to come from active galactic nuclei (AGN) which are divided, according to the orientation of the jets with respect to the line of sight, into blazars and MAGN.
Blazars are the most numerous population in the {\it Fermi}-LAT catalogs and are divided into BL Lacertae (BL Lac) objects and flat-spectrum radio quasars (FSRQs) according to the absence or presence of strong broad emission lines in their optical/UV spectrum, respectively.
The $\gamma$-ray emission from unresolved blazars has been estimated to be around 20-30\% of the integrated IGRB above 100 MeV (see e.g. \cite{DiMauro:2013zfa,Ajello:2013lka,Ajello:2011zi}.
MAGN are only a dozen in the $\gamma$-ray catalogs but the unresolved counterparts are expected to be much more numerous than blazars.
The diffuse $\gamma$-ray emission from MAGN is expected to be aroud 20-30\% of the integrated IGRB above 100 MeV \cite{2011ApJ...733...66I,DiMauro:2013xta}.
Finally, SF galaxies are predicted to have a numerous unresolved population and their contribution has been derived to be from a few \% up to almost the totality of the IGRB (see e.g. \cite{Ackermann:2012vca,Tamborra:2014xia}).
In particular the {\it Fermi}-LAT Collaboration \cite{Ackermann:2012vca} has deduced that the contribution of unresolved SF galaxies is among 4\% and 24\% of the integrated IGRB above 100 MeV and they have considered two different Spectral Energy Distribution (SED): a power law (PL) and a Milky Way (MW) function.
\\
Truly diffuse processes, like ultra-high energy cosmic-ray interactions with the extragalactic background light or intergalactic shocks, are other possible $\gamma$-ray emission mechanisms (see Ref.~\cite{2012PhRvD..85b3004C} and references therein).  
However, there are still quite large uncertainties associated to these processes and a subdominant contribution to the IGRB is theoretically possible (see e.g. Refs. \cite{2011PhLB..695...13B,2004APh....20..579G,2014arXiv1410.8697Z}).
Therefore, we do not take into account these contributions in the rest of the paper.
\\
We consider in our analysis the $\gamma$-ray emission from unresolved SF galaxy, MAGN, BL Lac, and FSRQ populations as derived in \cite{DiMauro:2013zfa,DiMauro:2013xta,Ajello:2011zi,Ackermann:2012vca}.
\begin{figure*}
\centering
\includegraphics[width=0.4\textwidth]{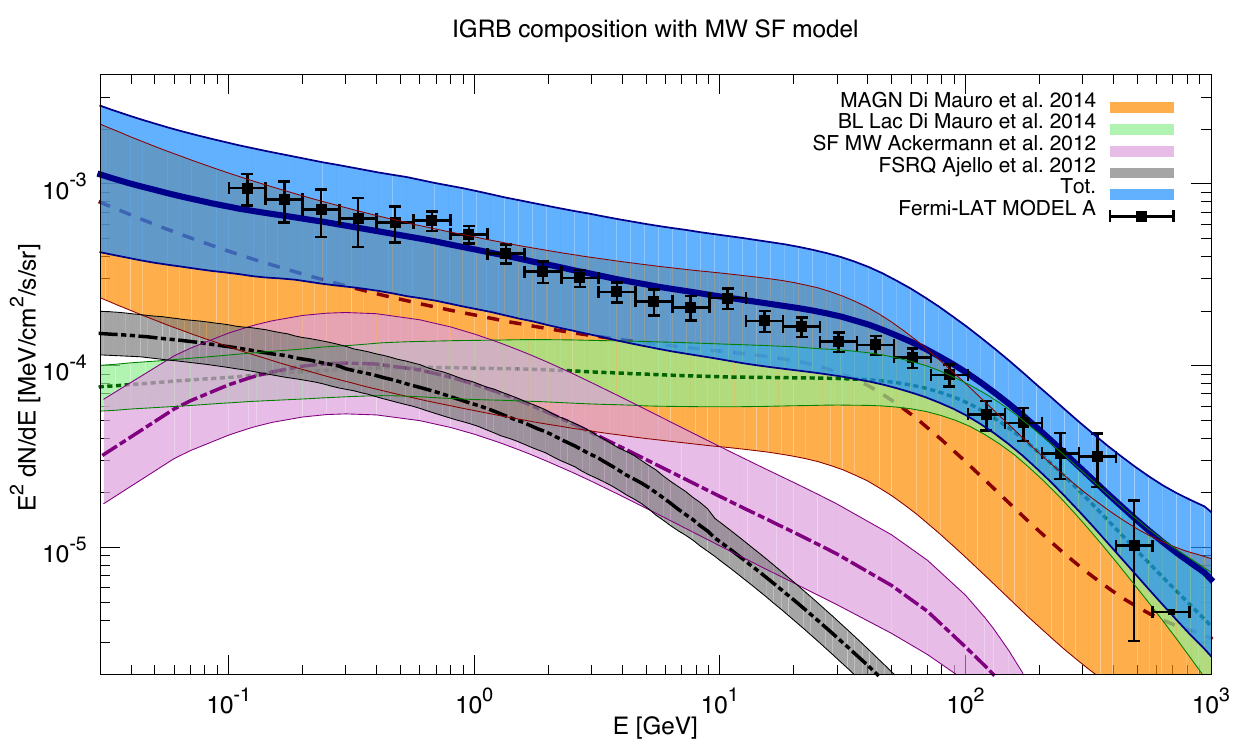}
\includegraphics[width=0.4\textwidth]{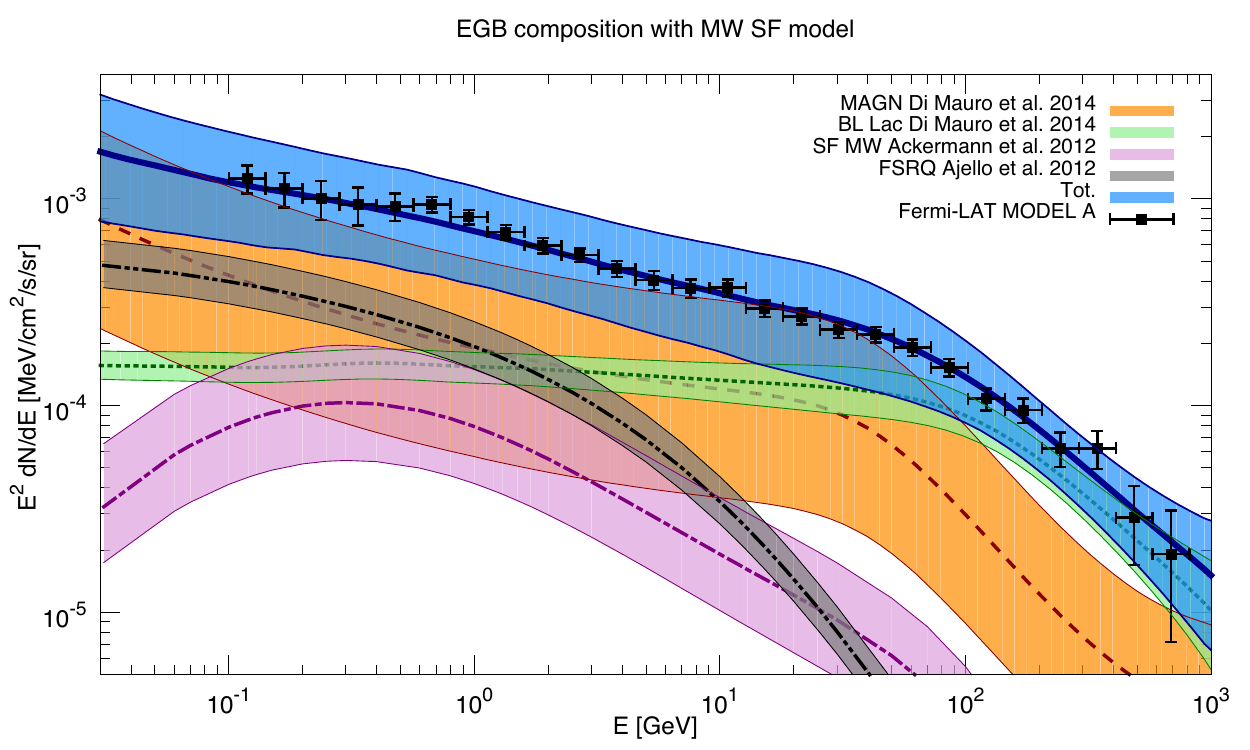}
\caption{Left (right) panels show the $\gamma$-ray emission from unresolved (unresolved+resolved) sources, together with the IGRB (EGB) data \cite{igrb_2014}. Lines and relevant uncertainty bands represent the contribution from the following source populations: 
MAGN (orange dashed),  BL Lacs (green dotted), FSRQs (grey double dot-dashed),  MW model of SF galaxies (purple dot-dashed), and the sum of all the contributions (blue solid). IGRB and EGB data have been derived with Galactic foreground Model A. }
\label{fig:igrb_egb_theo}
\medskip
\end{figure*}
In Fig.~\ref{fig:igrb_egb_theo} we display the {\it Fermi}-LAT IGRB and EGB data derived with the Model A of GDE \cite{igrb_2014}, together with the $\gamma$-ray fluxes predicted for AGN and SF galaxies. 
It is evident that both the IGRB and EGB data are consistent with the emission from these extragalactic populations.

We now determine with a chi-square statistical method at which extent AGN and SF galaxies can explain the IGRB and EGB data.
We perform a fit to the IGRB and EGB measurements with a $\chi^2$ function including both the errors of the {\it Fermi}-LAT data and the theoretical uncertainties of the $\gamma$-ray emission from astrophysical sources (see for all the details \cite{DiMauro:2015tfa}).
The theoretical uncertainties on AGN mainly produce a renormalization of their average unresolved $\gamma$-ray flux while the SF galaxy contribution contains also a large uncertainty due to the SED.
We thus consider, at the first order, the $1$-$\sigma$ error on the AGN and SF galaxy unresolved emission as the width of the bands shown in Fig.~\ref{fig:igrb_egb_theo} while the average fluxes are represented by the curves in Fig.~\ref{fig:igrb_egb_theo}.
Both the SF galaxy SED, namely the MW and the PL models, are taken into account to include also our ignorance on the spectral shape of these sources.
\\
We explore the possibility that the theoretical predictions adopted may be affected by an additional uncertainty on the spectral shape.
This possibility is included by varying the power-law index of AGN taking into account the relevant 1-$\sigma$ error on this parameter (see for further details \cite{DiMauro:2015tfa}).
The uncertainties on the SF galaxies SED is already considered by performing all our analysis with both the MW and PL models.
\\
The results are shown in Tab.~\ref{tab:fitbackgammafree}, where we display the reduced chi-square ($\tilde{\chi}^2 = \chi^2/{\rm d.o.f.}$) for the fit on the IGRB and EGB data.
The analysis is performed on the {\it Fermi}-LAT data derived with each of the benchmark GDE models considered in \cite{igrb_2014} and labelled with A, B and C. 
The $\gamma$-ray emissions from unresolved BL Lacs, FSRQs, MAGN and SF galaxies are able to explain the high-latitude IGRB and EGB data with no need for significant adjustments of the average parameters. 
A better agreement is provided by the MW modeling of the SF galaxy emission and with the Model C of the GDE \cite{igrb_2014}.
Indeed, the PL model of SF galaxies is in some tension with the data sets obtained with GDE Model A and B  (see Tab.~\ref{tab:fitbackgammafree}).  
The choice of the Galactic foreground model has a large relevance on the goodness of the fit. 
\begin{figure*}
\centering
\includegraphics[width=0.4\textwidth]{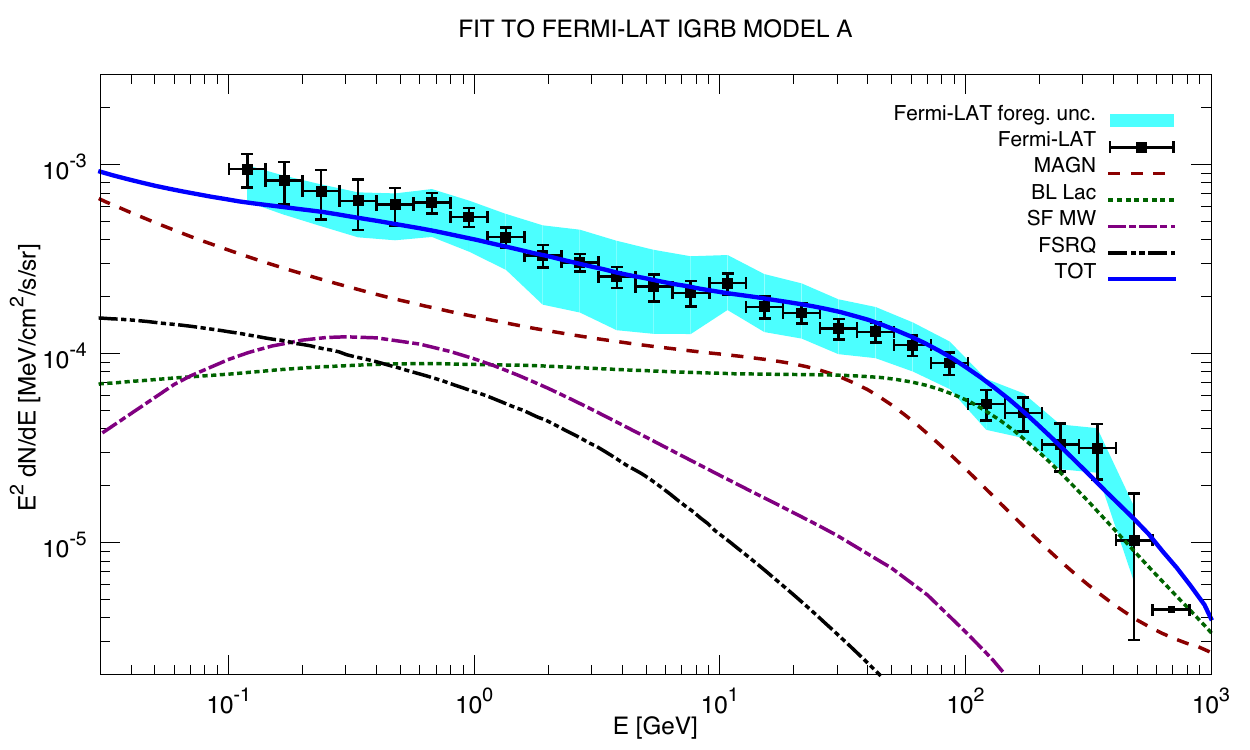}
\includegraphics[width=0.4\textwidth]{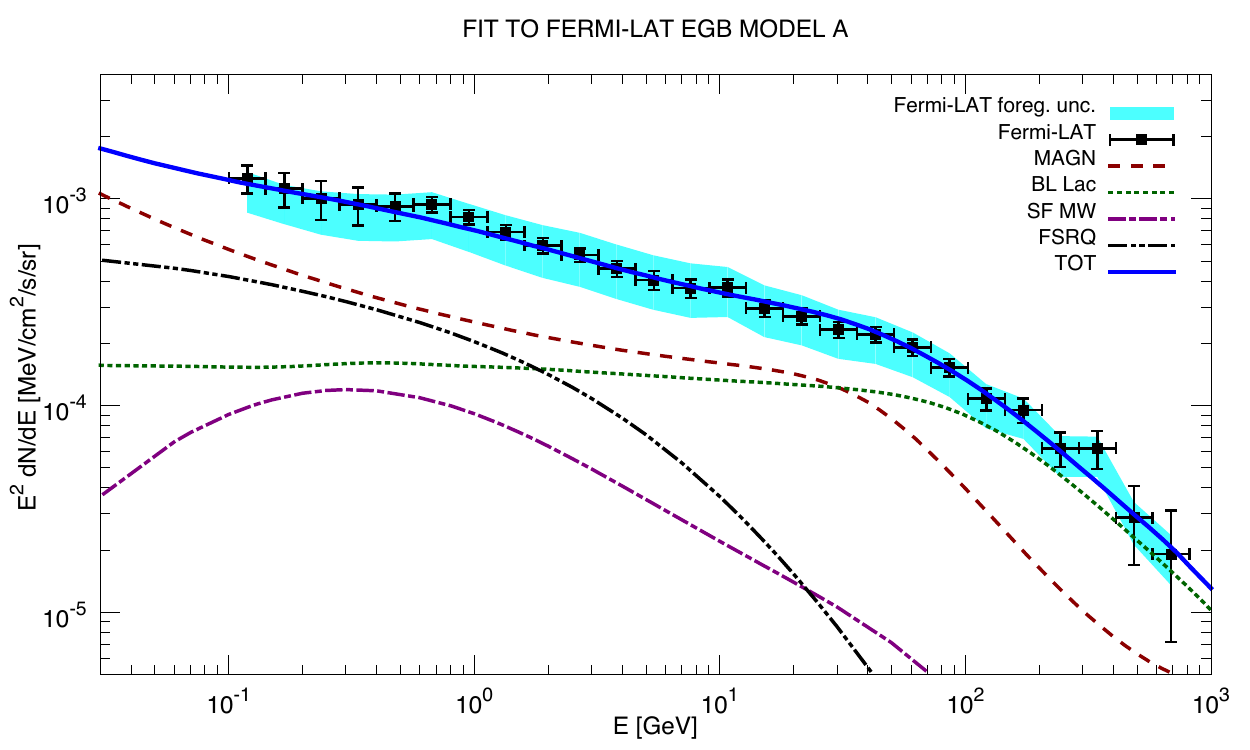}
\caption{The $\gamma$-ray flux best fit on Model A of IGRB (left panels) and EGB (right panels) data is shown for: BL Lac (dotted green), MAGN (dashed brown), SF (dot-dashed purple), FSRQ population (dot-dot-dashed black), the sum of AGN and SF (solid blue). 
We display with a cyan band also the uncertainty associated to the {\it Fermi}-LAT GDE model \cite{igrb_2014}.}
\label{fig:backfit}
\medskip
\end{figure*}
We display in Fig.~\ref{fig:backfit} the fluxes corresponding to the best fit configuration with the IGRB 
and the EGB data obtained with Model A of the GDE. 
It is remarkable that each contribution has a very different shape but they add in a way that their sum is consistent with a power-law with an exponential cutoff.
\begin{table}[t]
\center
\scalebox{0.9}{
\begin{tabular}{|c|c|c|c|c|}
\hline
 $\tilde{\chi}^2$  &  IGRB (MW)  &   EGB (MW)    &  IGRB (PL)    &  EGB (PL)   \\
 \hline
MODEL A  &   1.72; 1.56   &   0.95; 1.02     &  3.20; 2.54    &   1.41; 1.36    \\
MODEL B  &   1.33; 1.32   &   1.57; 1.72     &  2.30; 1.96    &   1.83; 2.06    \\
MODEL C  &   0.82; 0.84   &   0.60; 0.60     &  1.67; 0.95    &   0.77; 0.84    \\ 
\hline
\end{tabular}
}
\caption{The reduced chi-square value $\tilde{\chi}^2 = \chi^2/{\rm d.o.f.}$ for the fits performed using only the normalizations as free parameters (left numbers in each column) or varying also the  slope of the spectra (right numbers in each column). 
The analysis is performed with both the MW and PL SF galaxy models and for the three Galactic foreground models of the IGRB and EGB data \cite{igrb_2014}.}
\label{tab:fitbackgammafree}
\end{table}


\section{Constraints on a DM contribution to the {\it Fermi}-LAT IGRB data}
\label{sect:dm}
A possible contribution to the high latitude $\gamma$-ray IGRB could arise from annihilating DM particles present both in the halo of the MW and in external halos \cite{Calore:2014hna}.
DM can produce $\gamma$ rays both directly (the so-called {\it prompt} emission), or indirectly via the inverse Compton scattering of the electrons and positrons, produced by the DM annihilation, off the ambient light of the interstellar radiation field.
In order to simplify the discussion we consider only the DM distributed in a halo of the MW.
Moreover, we do not consider any specific particle physics model and we fix the branching ratio equal to 1 for any of the discussed annihilation channels.
The photon and electron spectra have been calculated using the Pythia Montecarlo code for DM annihilations into $e^{+}e^{-}$, $\mu^{+}\mu^{-}$, $\tau^{+}\tau^{-}$, $b\bar{b}$, $t\bar{t}$ and $W^+W^{-}$ 
channels.
We have assumed an Einasto DM profile with a local DM density of $\rho_{\odot}=0.4$ GeV/cm$^3$.
For all the details about the $\gamma$-ray flux due to DM annihilation in the halo of the MW we refer to \cite{Calore:2013yia,DiMauro:2015tfa}.

\begin{figure*}
\centering
\includegraphics[width=0.4\textwidth]{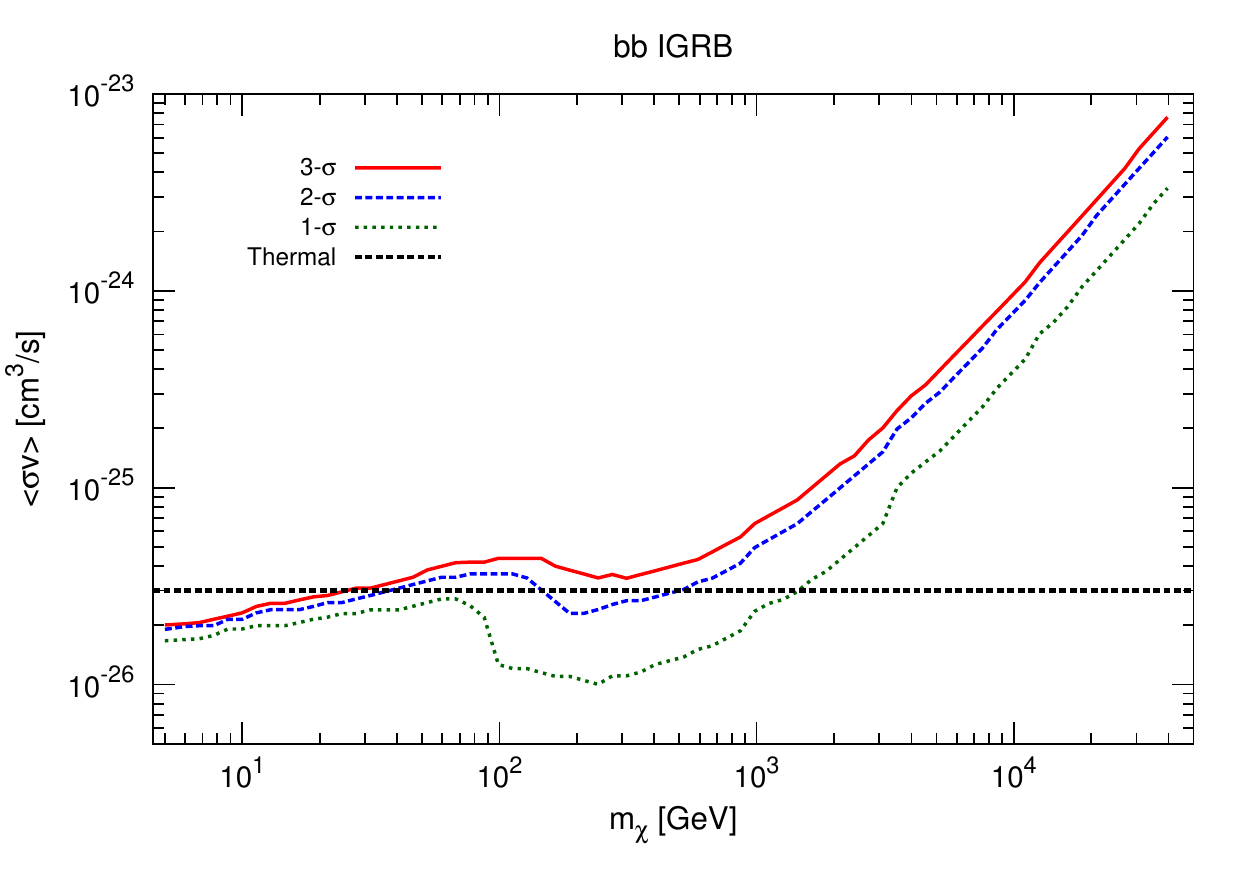}
\includegraphics[width=0.4\textwidth]{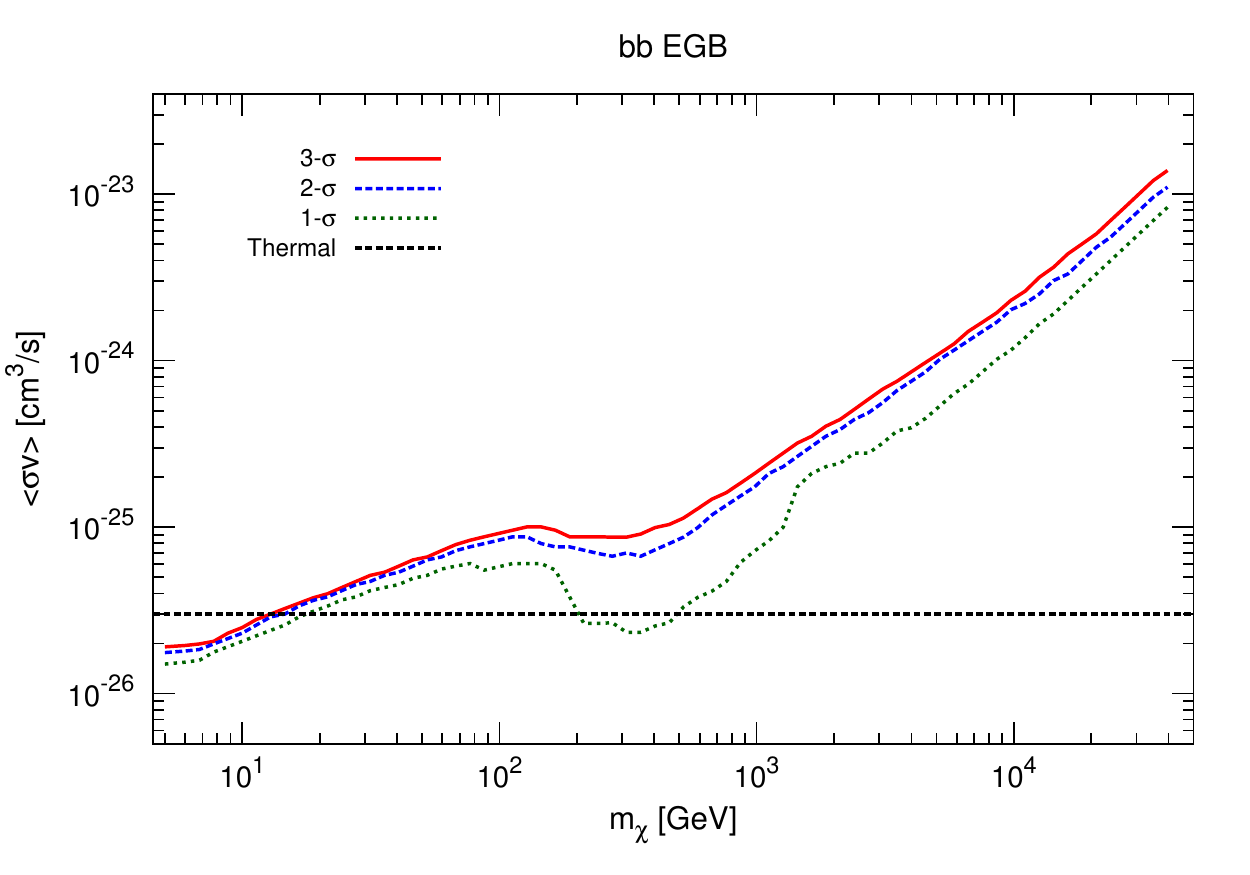}
\includegraphics[width=0.4\textwidth]{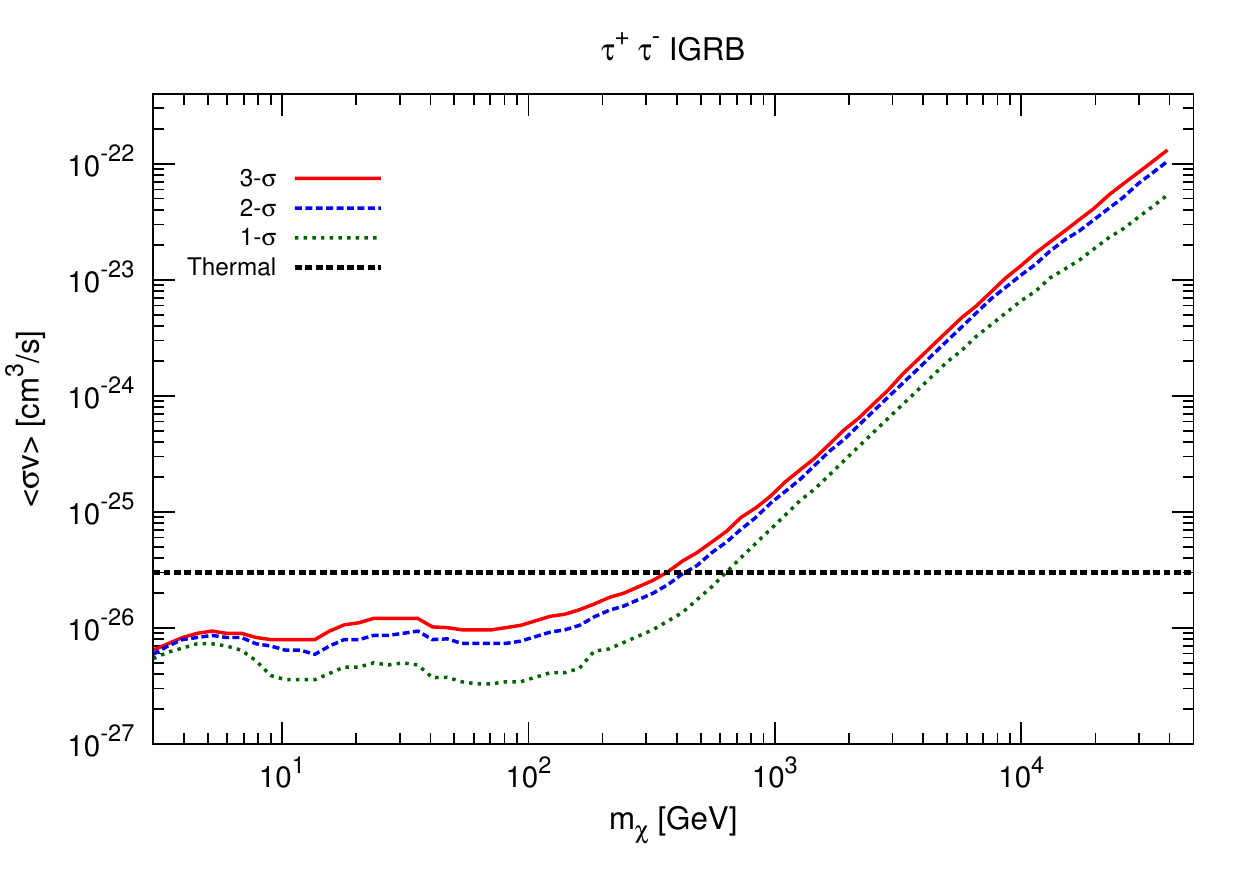}
\includegraphics[width=0.4\textwidth]{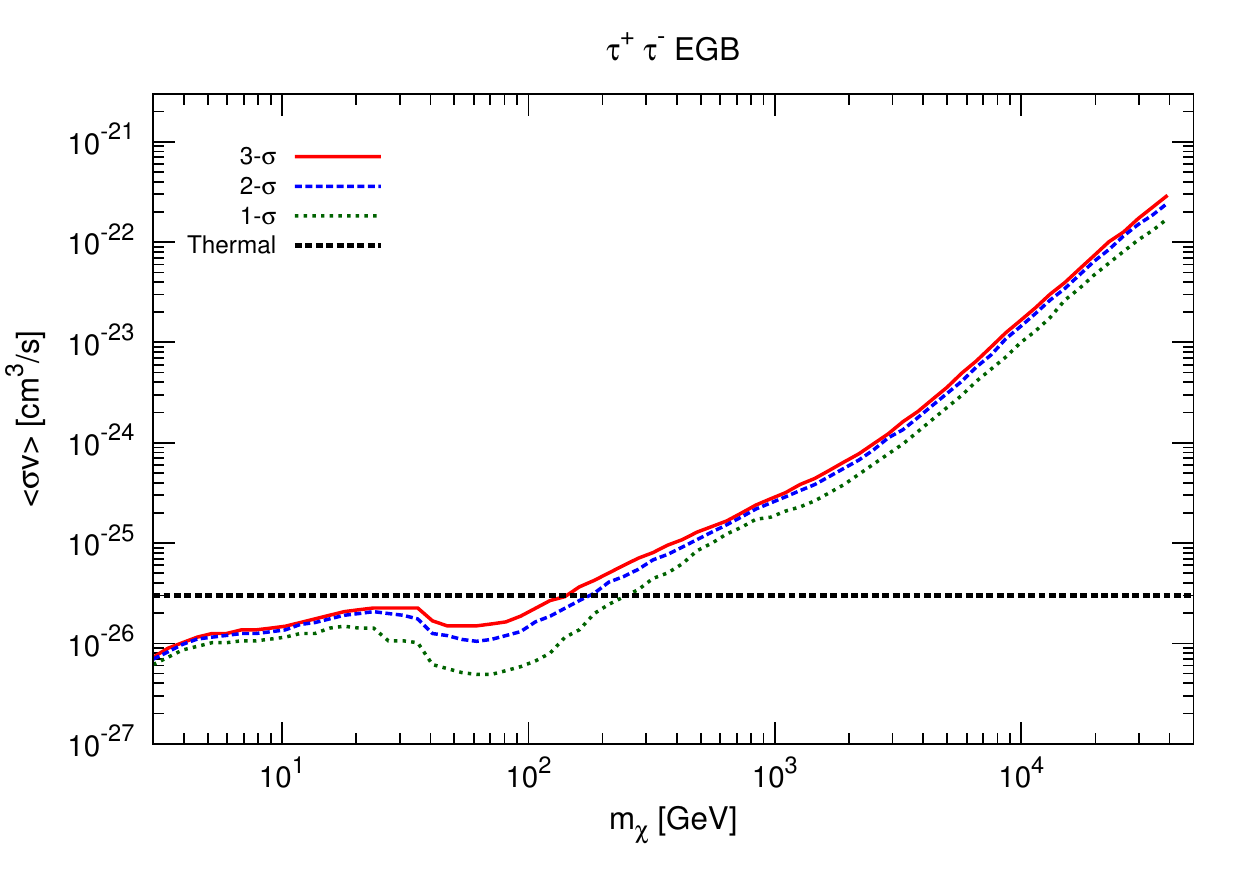}
\caption{Upper limits on \sigmav as a function of the DM mass $m_\chi$ for $b\bar{b}$ (top panels) and $\tau^+\tau^-$ (botton panels) channels. The upper limits are derived with a fit to the IGRB (left panels) and EGB (right panels) data, within GDE Model A. The 3-$\sigma$, 2-$\sigma$ and 1-$\sigma$ C.L.s are shown with solid, dashed and dotted curves, respectively. The horizontal dotted line specifies the thermal relic annihilation cross section value.}
\label{fig:UL_IGRB}
\medskip
\end{figure*}

As a first analysis, we derive upper limits for the DM annihilation cross section $\langle\sigma v\rangle$, fitting the IGRB and EGB data with the astrophysical sources discussed in Sect.~\ref{sec:astro} and the addition of a Galactic DM halo.
The results are shown in Fig.~\ref{fig:UL_IGRB} for the DM annihilation channels listed above, different 
confidence levels (C.L.s) and considering the Model A of the Galactic foreground. 
In the case of the $b\bar{b}$ DM annihilation channel and for masses lighter than 
30 GeV, the upper bounds on \sigmav are below the thermal relic value while at 10 TeV our analysis excludes $\langle\sigma v\rangle \gsim 10^{-24} {\rm cm}^3 / {\rm s}$. 
Moreover, the limits obtained for the DM annihilation into $\tau^+\tau^-$ are quite stringent: the thermal relic cross section is excluded at 3-$\sigma$ C.L. up to a DM mass of about 330 GeV, while at $m_\chi \simeq1$ TeV the bound is around $10^{-25} {\rm cm}^3 / {\rm s}$. 
As expected, the upper bounds obtained with a fit to the EGB data are very similar to the ones obtained with the IGRB data and in general are only slightly looser.  
\begin{figure*}
\centering
\includegraphics[width=0.4\textwidth]{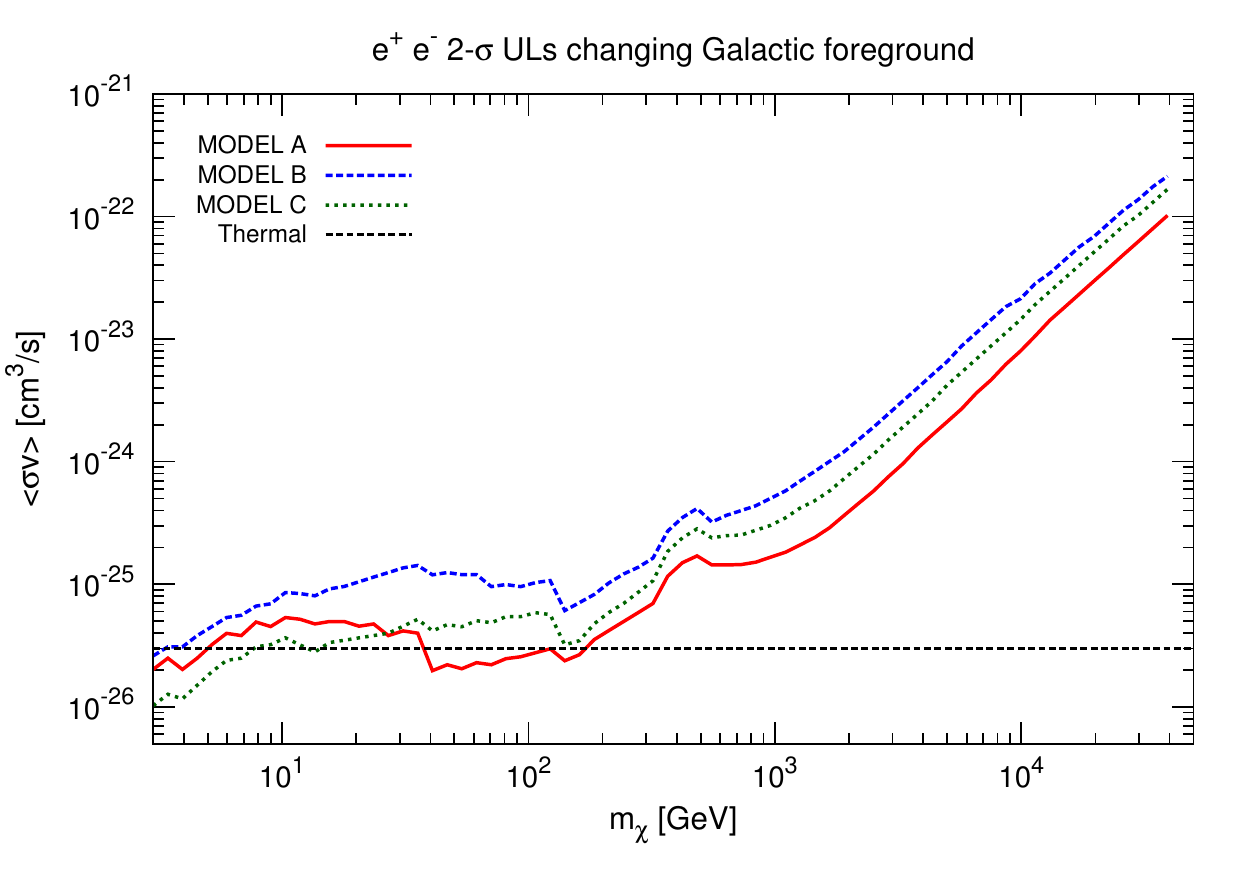}
\includegraphics[width=0.4\textwidth]{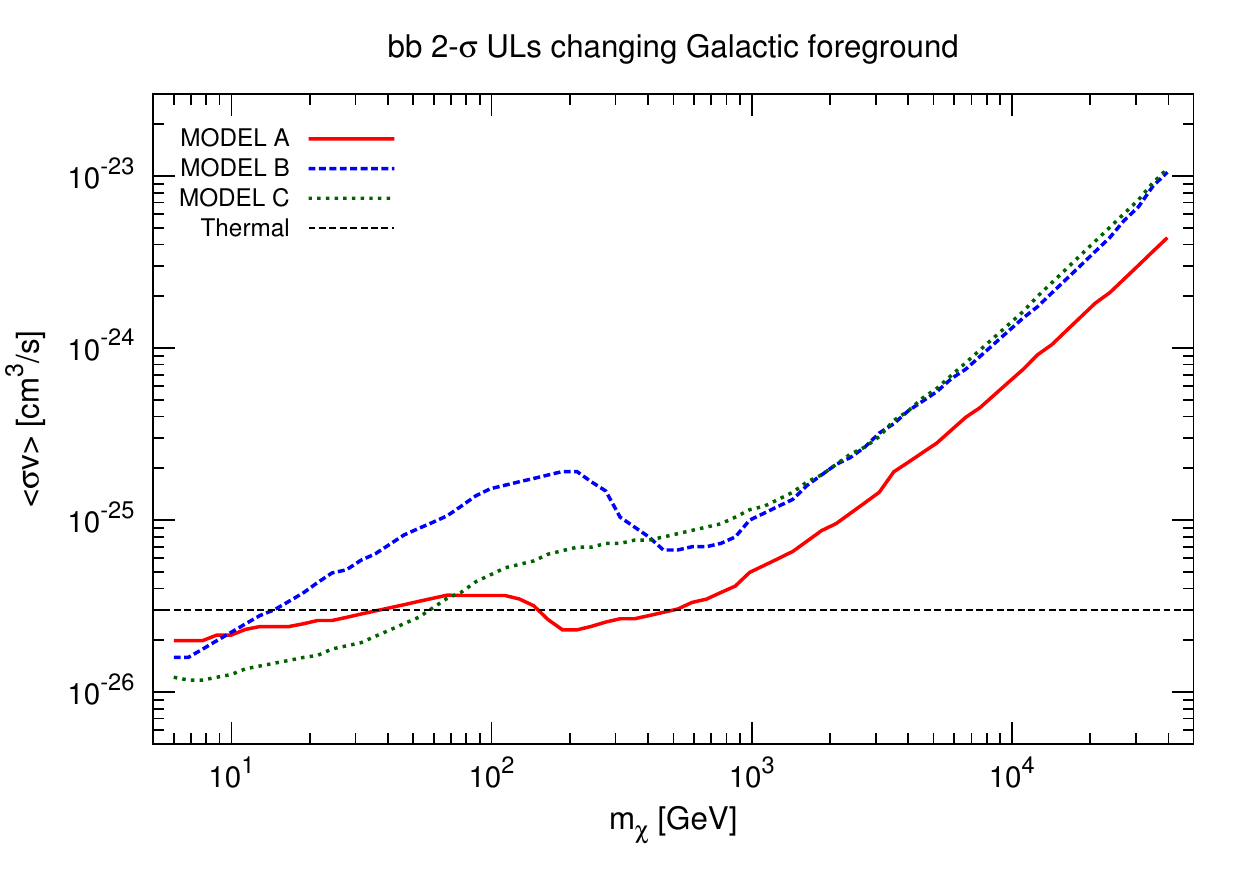}
\caption{2-$\sigma$ C.L. upper limits on the annihilation cross section for the $e^+e^-$ (left panels) and $b\bar{b}$ (right panels) channels obtained with the IGRB data derived with three different GDE models \cite{igrb_2014}.
}
\label{fig:gal_foreground}
\medskip
\end{figure*}
\\
For the first time we explore the impact of the GDE model in the estimation of upper limits on the DM annihilation cross section. 
The results are illustrated, for a 2-$\sigma$ C.L and for the annihilation channels $e^+e^-$ and $b\bar{b}$, in Fig.~\ref{fig:gal_foreground}.
The choice of the GDE model turns out to have a significant 
role in the values of the upper limits on $\langle\sigma v\rangle$. 
The results vary on average within a factor of two but the differences can reach a factor of 10 in the case of 
$b\bar{b}$ at about $m_\chi \simeq 200$ GeV.
The bounds obtained on the IGRB with Model B of the GDE are always looser with respect to the ones derived with Model A or C. 
\begin{figure*}
\centering
\includegraphics[width=0.32\textwidth]{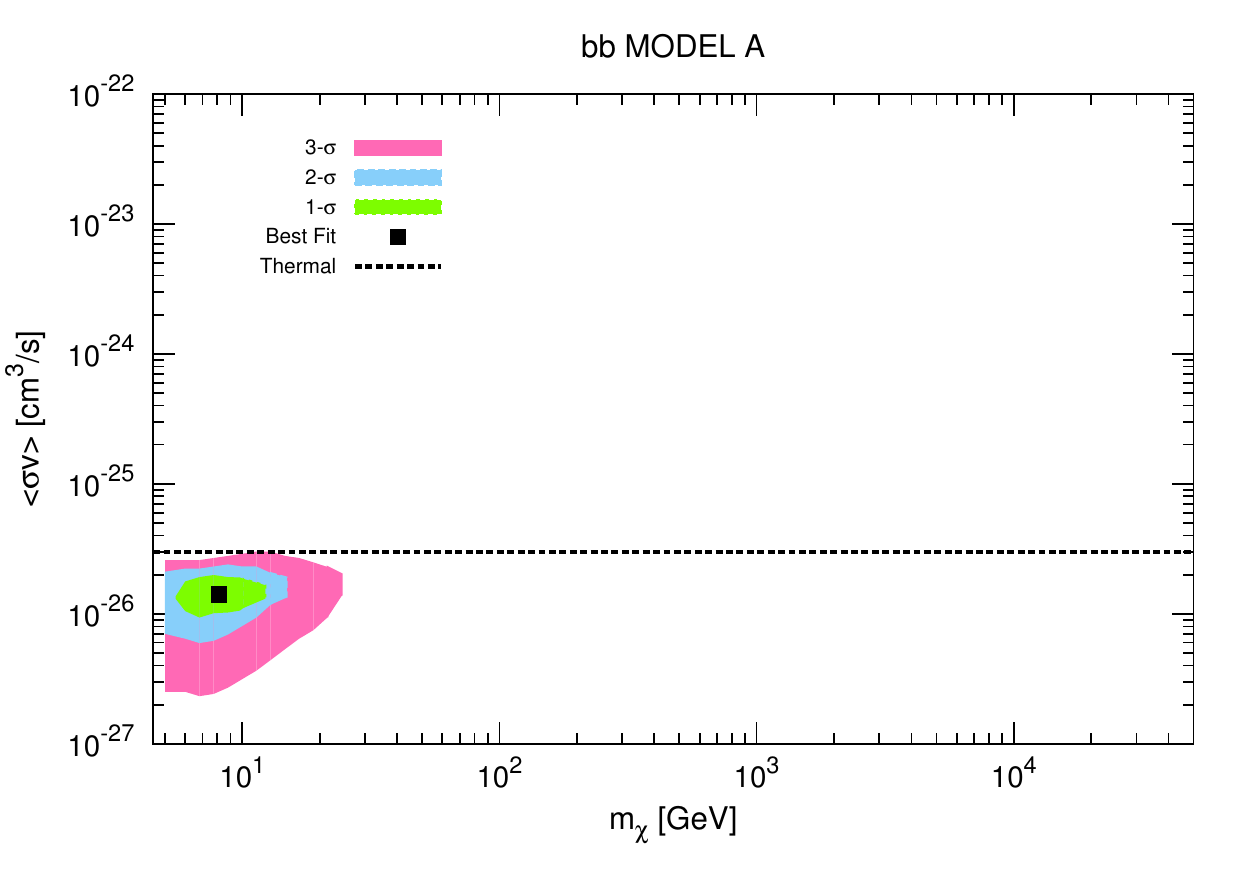}
\includegraphics[width=0.32\textwidth]{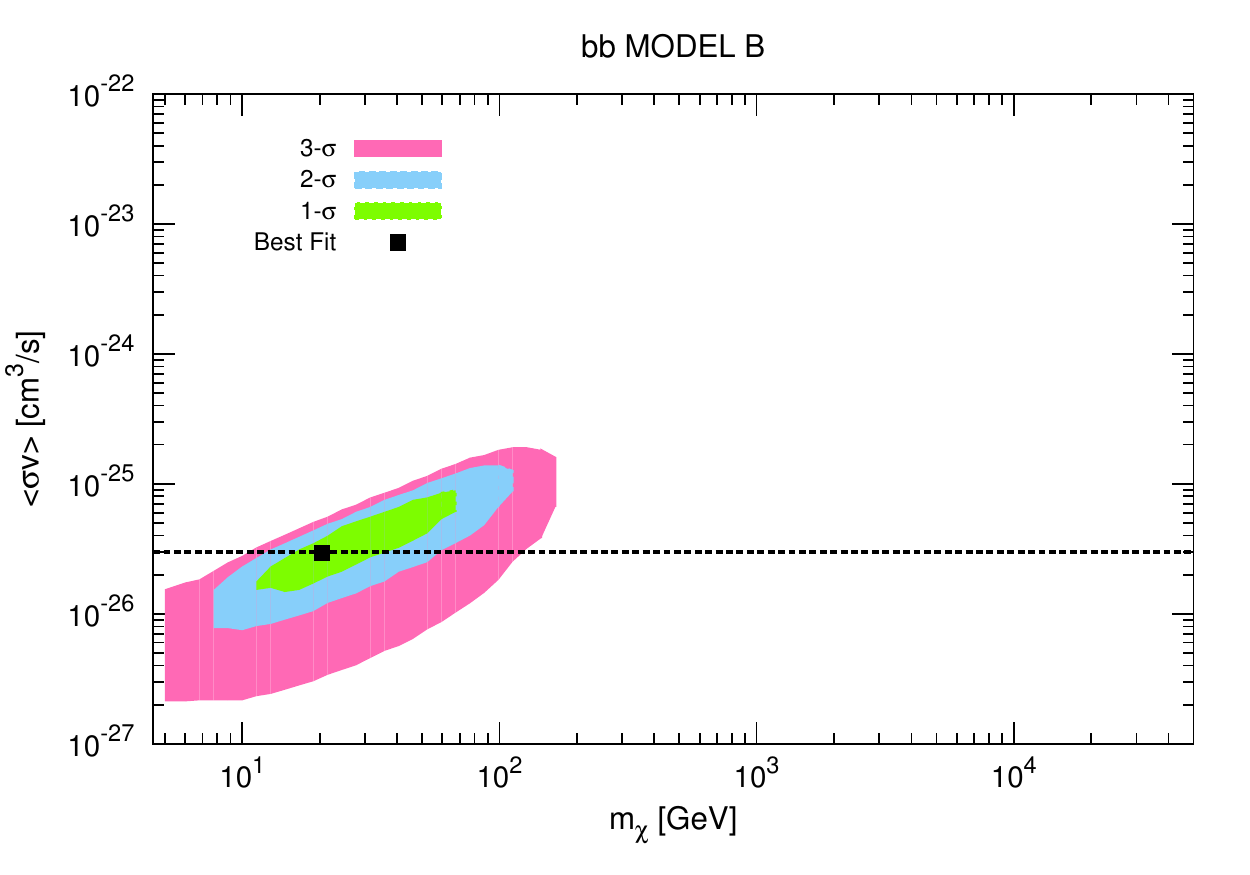}
\includegraphics[width=0.32\textwidth]{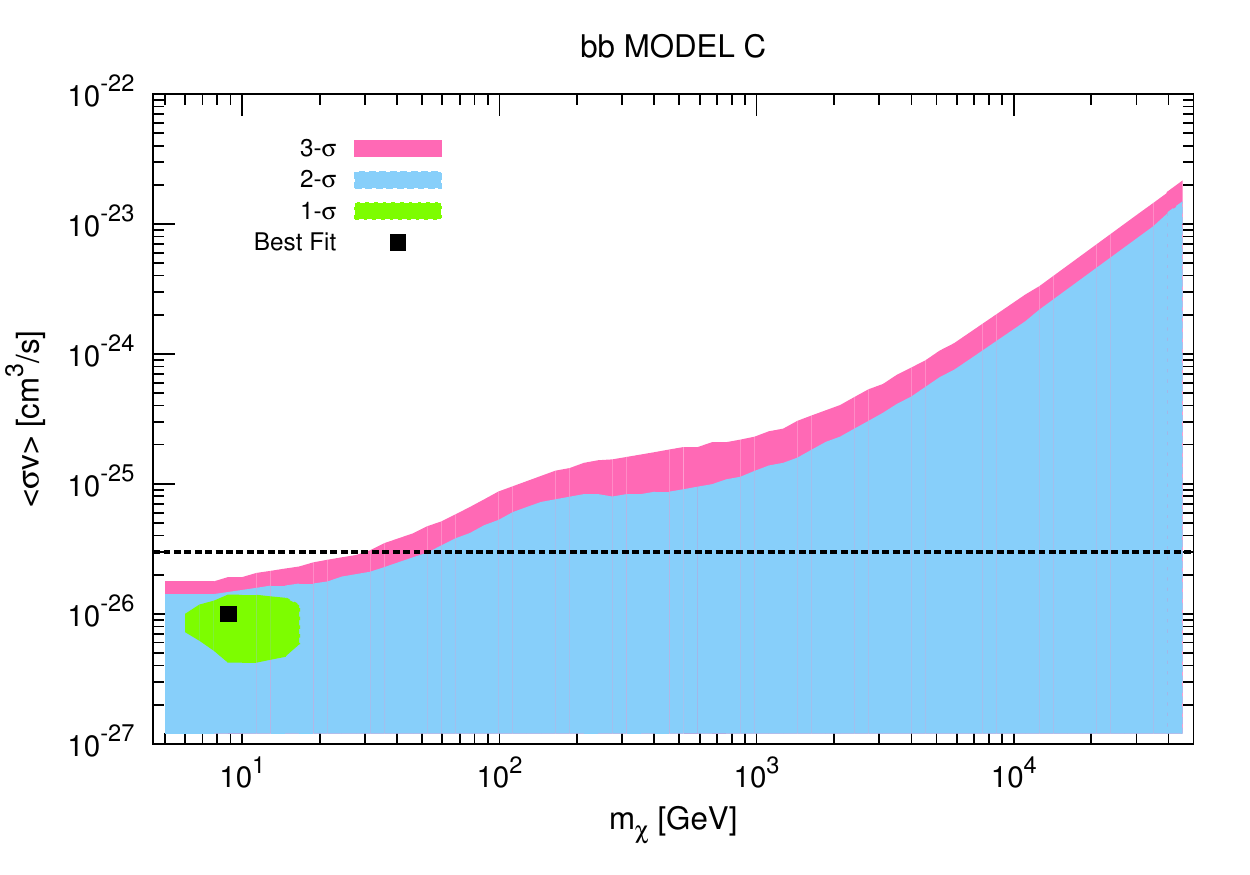}
\caption{Contour plots in the $\langle\sigma v\rangle$-$m_\chi$ plane for $b \bar{b}$ channel calculated with a fit of the astrophysical sources and a DM component on the IGRB data. The black dots refer to the values of the best fit, and the (closed or open) green, blue and pink regions indicate the 1, 2 and 3-$\sigma$ C.L. Left, center and right panels correspond to the case of Model A, B and C of the IGRB.}
\label{fig:cp_igrb}
\medskip
\end{figure*}

As a second analysis, we attempt to identify DM configurations which can significantly improve the fit to the IGRB data. 
In this part we perform a fit to the {\it Fermi}-LAT data with extragalactic sources and a DM component 
as done before, but letting the WIMP DM mass $m_\chi$ and  \sigmav varying simultaneously. 
Our results are displayed in Fig.~\ref{fig:cp_igrb}, for 
the $b \bar{b}$ DM annihilation channel and for the IGRB data associated to the three Galactic foreground  models.   
In the case of Model A and B, we obtain closed regions up to 3-$\sigma$ C.L. with the best fits located around $m_\chi \simeq$ 5-20 GeV and  \sigmav $\simeq 1$-$3\cdot 10^{-26}\rm{cm}^3/\rm{s}$ while for Model C the 1-$\sigma$ C.L. closed region opens up at already 2-$\sigma$ C.L., translating the results into upper limits. 
The addition of a DM component is almost irrelevant for IGRB Model C while, in the case of leptonic or $b \bar{b}$ channels, improves the IGRB fit, for models A and B, with $\Delta \chi^2 = \chi^2_{\rm{astro}} - \chi^2_{\rm{astro}+\rm{DM}} \geq 6.5$, where $\chi^2_{\rm{astro}}$ and $\chi^2_{\rm{astro}+\rm{DM}}$ are the best fit $\chi^2$ with only astrophysical sources or with the addition of a DM component. 
In the case of $e^+e^-$, $\tau^+\tau^-$ and $b\bar{b}$ channels, the best fit values of the DM mass range from few GeV up to 20 GeV and the annihilation cross section values are close to the thermal one while the 
$\mu^+\mu^-$ channel requires \sigmav $\simeq 1$-$3\cdot 10^{-25}\rm{cm}^3/\rm{s}$. 
It is remarkable that including a DM component does not require the standard astrophysical contributions to differ significantly from the average emission \cite{DiMauro:2015tfa}. 
Therefore, a DM component can very well fit the IGRB data with a realistic unresolved emission from extragalactic sources. 

The results illustrated in Fig.~\ref{fig:cp_igrb} demonstrate that a DM contribution to IGRB may significantly improve the fit to the IGRB with respect to the interpretation with only astrophysical source populations.
However, the significance of this potential exotic signal strongly depends on the choice of the Galactic foreground model considered to derive the IGRB spectrum.
It is evident in Fig.~\ref{fig:cp_igrb} that for Model A and B of the GDE we may have an hint of DM up to 3-$\sigma$ C.L. while for Model C for 2 and 3-$\sigma$ C.L. we can only set upper limits on the  annihilation cross section.
These results confirm how a deep knowledge of the GDE is mandatory to unveil a DM contribution in the IGRB data. 
\begin{figure*}
\centering
\includegraphics[width=0.43\textwidth]{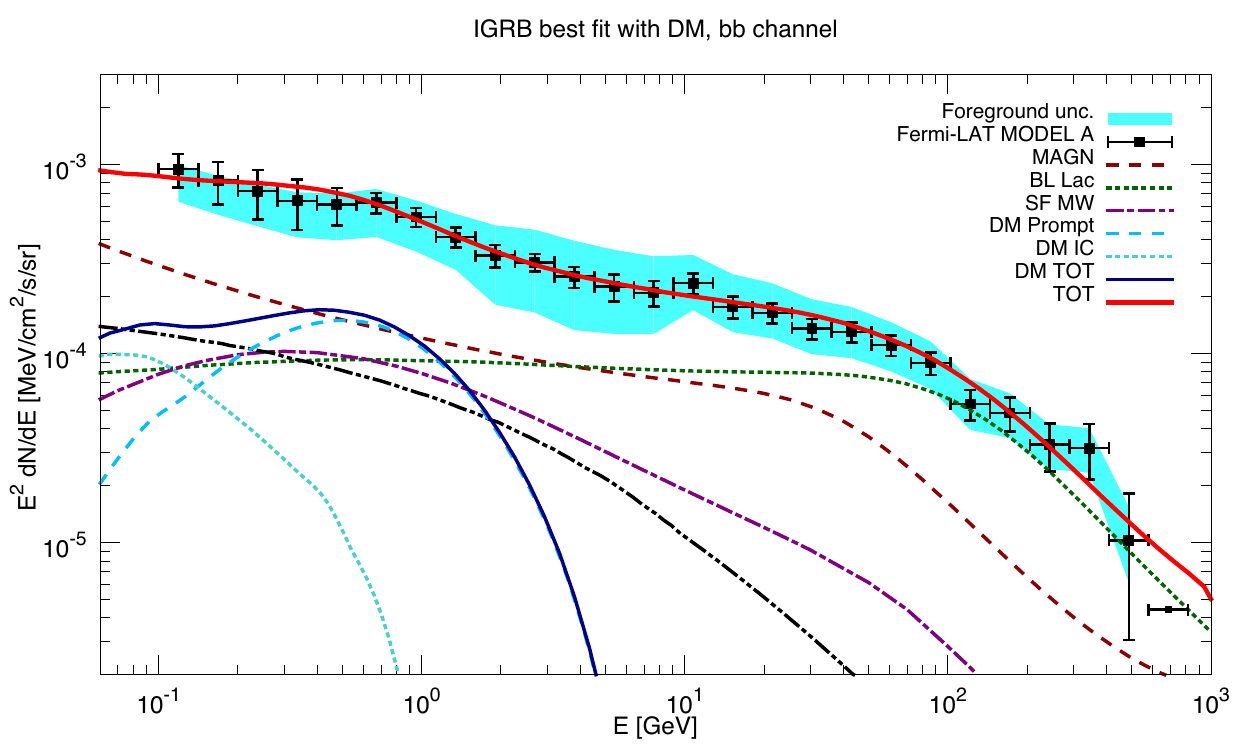}
\includegraphics[width=0.43\textwidth]{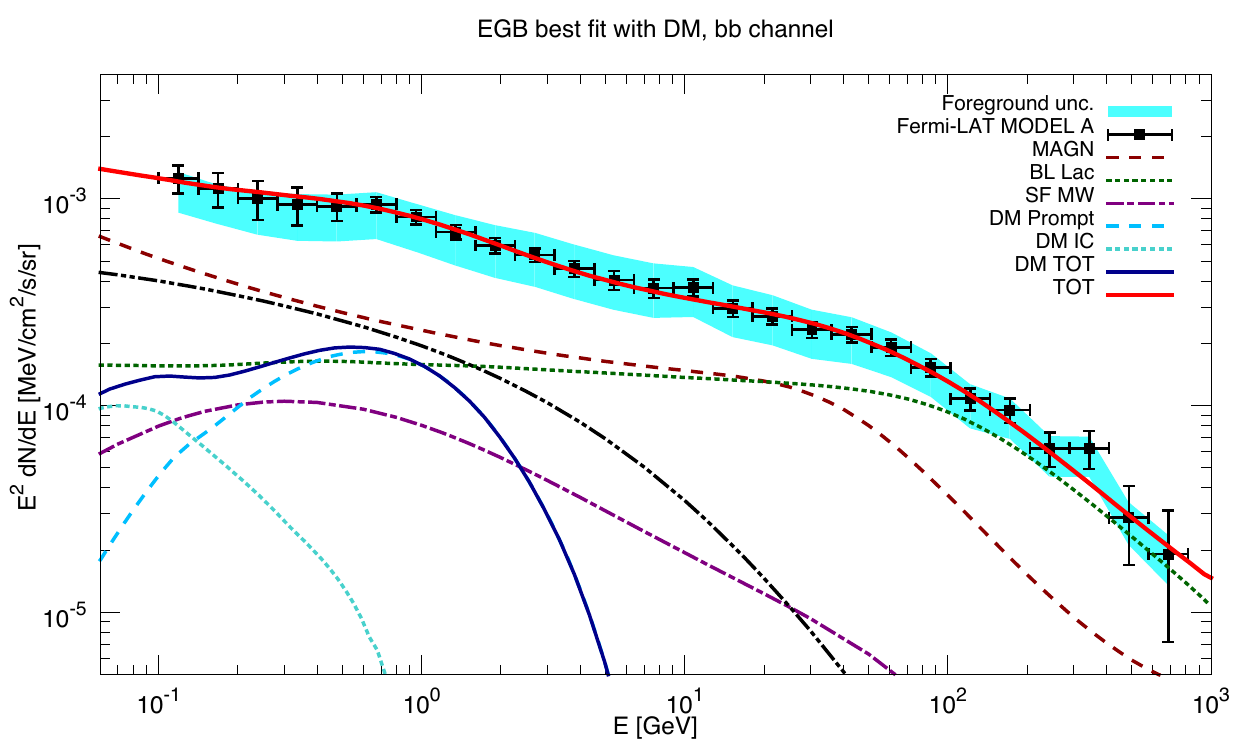}
\caption{The differential $\gamma$-ray flux for the unresolved (left panels) and entire (right panels) BL Lac, FSRQ, MAGN, SF galaxy populations and the DM contribution with $b\bar{b}$ channel (splitted into the prompt and the ICS emission) as fixed by the best fit to the Model A of the IGRB and EGB data.}
\label{fig:DM_bestfit}
\medskip
\end{figure*}
In Fig.~\ref{fig:DM_bestfit} we illustrate the best fit configuration for the $b\bar{b}$ DM annihilation channel and Moled A of the GDE.
This is the case with the largest significance of DM with respect to the best fit with only astrophysical populations with $\Delta \chi^2 = 18.9$.
In this specific best fit, the DM mass is $m_\chi$=8.2 GeV, and the annihilation cross section is $\langle\sigma v\rangle =1.4 \cdot 10^{-26}\rm{cm}^3/\rm{s}$. 
It is clear in Fig.~\ref{fig:DM_bestfit} that this configuration of extragalactic sources and DM reproduce very well the IGRB and EGB data.

The upper bounds reported in Figs.~\ref{fig:UL_IGRB} and \ref{fig:gal_foreground} improve the results derived in \cite{Calore:2013yia}, in the so-called `best-fit' scenario, by a factor of $\sim$3 at $m_{\chi} \sim 10$ GeV and a factor of 30 at $m_{\chi} \sim 10$ TeV.
Our limits also improve significantly the upper limits on \sigmav derived by the {\it Fermi}-LAT analysis for a Galactic halo of DM \cite{Ackermann:2012rg} and with the analysis of 25 dwarf Spheroidal galaxies \cite{Ackermann:2013yva}.
We notice that for $m_\chi\simeq$ 10 TeV the upper bounds found in \cite{Abramowski:2014tra} by the H.E.S.S. Collaboration, which are optimized at energies larger than about 1 TeV, are of the same entity as ours for the leptonic channels, while for hadronic channels they are about one order of magnitude weaker.
In addition, we derive similar results with respect to the very recent analysis performed by the {\it Fermi}-LAT Collaboration in Refs. \cite{Ajello:2015mfa,Ackermann:2015tah}.

\section{Conclusions}
We have performed the first detailed statistical analysis for the interpretation of the recent IGRB data, measured by the {\it Fermi}-LAT Collaboration \cite{igrb_2014}. 
We first test the hypothesis that a numerous sample of unresolved extragalactic sources may explain the {\it Fermi}-LAT IGRB data.
For the first time a $\chi^2$ function which includes the data errors and the theoretical uncertainties on the $\gamma$-ray emission from BL Lacs, FSRQs, MAGN and SF galaxies has been considered.
The theoretical uncertainty from each of this extragalactic population has been first parametrized only with a renormalization factor and then generalized adding also a possible change in the photon index of the $\gamma$-ray SED. Since the spectral shape of SF galaxies is not well constrained we have considered both the MW and the PL models.
In our results the IGRB and the EGB data are well fitted by the unresolved emission from AGN and SF galaxies with best fit parameters close to the average theoretical values.
We also demonstrate how the choice of the Galactic foreground model, used to derive the IGRB and EGB data, affects the results.
\\
We explore also a possible contribution from the annihilation of WIMP DM particles distributed in the halo of our Galaxy. 
As a first analysis, we derive upper limits on the DM annihilation cross section, combining the $\gamma$-ray emission from astrophysical sources and DM in order to fit the IGRB and EGB data.
The upper bounds calculated with this method are stringent and rule out the thermal relic cross section for a wide range of DM mass values for the $b\bar{b}$ and $\tau^+ \tau^-$ annihilation channels. 
\\
We finally derive DM configurations which improve the fit to the IGRB data with respect to the case with only astrophysical sources.
The best fit DM mass ranges from few GeV up to 
20 GeV and the annihilation cross section $\langle\sigma v\rangle$ values are close to the thermal one. 
A DM component may fit, together with AGN and SF galaxies, very well the IGRB and EGB data with best fit parameters for the astrophysical populations close to the average theoretical values.
However depending on the Galactic foreground model and the value of the C.L., we obtain a possible hint of DM or we set only upper limits.
\\
Our results show how crucial is the IGRB in the study of the extragalactic source populations and for DM searches.
It is today a powerful tool to constrain the DM properties and with a future improvement in the knowledge of the GDE and of the unresolved emission from AGN and SF galaxies, may probe a DM contribution to the $\gamma$-ray sky.

\bibliographystyle{ieeetr} 
\bibliography{procdima}

\end{document}